\documentclass{iaus}
\usepackage{graphicx}

\title[PP ??.~~The SFH of fossil group galaxies] %% give here short title %%
{The star formation histories of fossil group galaxies}

\author[de la Rosa, Proctor, Mendes de Oliveira, Forbes, Cid Fernandes \& Mateus]   %% give here short author list %%
{Ignacio G. de la Rosa$^1$, Robert N. Proctor$^2$, Claudia Mendes de Oliveira$^2$, Duncan A. Forbes$^3$, Roberto Cid Fernandes$^4$  \and Abilio Mateus$^4$}  

\affiliation{$^1$Instituto de Astrof\'{i}sica de Canarias, Spain \\ email: {\tt irosa@iac.es} \\[\affilskip]
$^2$Instituto Astron\^{o}mico e Geof\'{i}sico-USP, Brazil \\[\affilskip] 
$^3$Centre for Astrophysics \& Supercomputing-SU, Australia \\[\affilskip]
$^4$Universidade Federal de Santa Catarina, Brazil 
}

\pubyear{2009}
\volume{262}  %% insert here IAU Symposium No.
\pagerange{1--2}
\jname{Stellar Populations - Planning for the Next Decade}
\editors{G.R. Bruzual \& S. Charlot, eds.}
\begin{document}

\maketitle

\begin{abstract}
A comparison is carried out among the star formation histories of early-type galaxies in fossil groups, clusters and low density environments. Although they show similar evolutionary histories, a significant fraction of the fossils are younger than their counterparts, suggesting that they can be precursors of the isolated ETG galaxies. 
\end{abstract}

\firstsection % if your document starts with a section,
\section{Introduction}

Low density and void environments are mainly populated by late-type galaxies. Nevertheless, the presence of isolated and quiescent early-type galaxies (ETGs) challenges the current galaxy formation ideas. Some authors propose an environment-independent mechanism, like radio-mode AGN heating, to quench the star formation in isolated galaxies (\cite[Croton \& Farrar 2008]{Croton08}). As the mechanism only depends on the host halo mass, similar evolutionary histories would be observed for galaxies in both low- and high-density environments. Contrarily, other authors propose that all bright galaxies in a very dense environment have merged into isolated ones (e.g. \cite[Balogh \etal\  2004]{Balogh04}). In this context, fossil groups fit the bill, as they are considered the 'final stage' of the merging of L* galaxies in groups or small clusters (e.g. \cite[Ponman \etal\ 1994]{Ponman_etal94}). In the present study, we test these two models by comparing the Star Formation Histories (SFH) of the first-ranked ellipticals in fossil groups and clusters to the SFH of their counterparts in low density environments.

\begin{figure}
\begin{center}
 \includegraphics[width=4.9in]{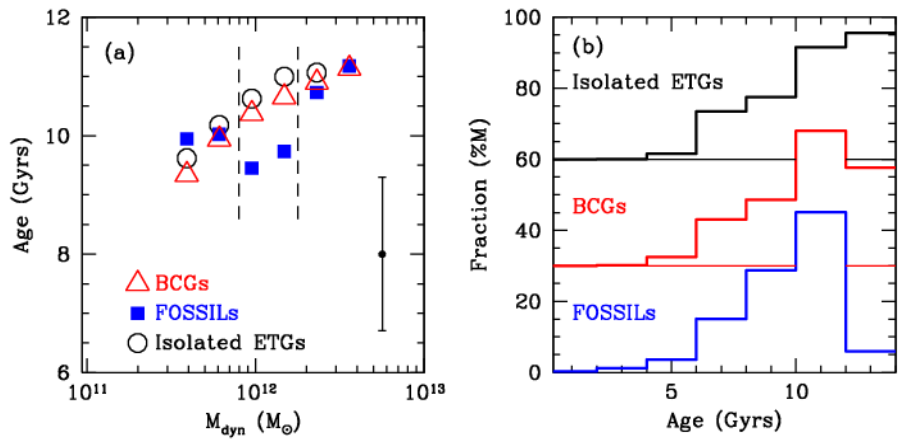} 
 \caption{(a) Mass-weighted mean ages for M$_{dyn}$-binned galaxies. Typical dispersion around the mean in the low-right corner. (b) Average SFH for the galaxies in the two central M$_{dyn}$ bins, displaced vertically for clarity}
   \label{fig1}
\end{center}
\end{figure}

\section{The data}

Our sample contains 9222 galaxies from the DR7-SDSS database, divided into three subsamples. The 'fossil subsample' with 35 systems, composed by a subset of the fossil groups in \cite[Santos \etal\ (2007)]{Santos07}, a few of the well known fossils from the list of \cite[Mendes de Oliveira \etal\ (2006)]{Mendes06} as well as some newly found ones. The other two subsamples were selected to match both the fossil's R$_{eff}$ (kpc) and M$_{dyn}$. The Bright Cluster Galaxy 'BCG subsample' with 7151 objects extracted from the MaxBCG Catalog (\cite[Koester \etal\ 2007]{Koester07}) and the 'Isolated ETG subsample' from the SDSS-DR7, with further constrains to restrict the sample to ETGs in low density environments.

\section{Results of the Stellar Population study}

The SFHs were obtained through spectral fitting with the STARLIGHT code (\cite[Cid Fernandes et al. 2005]{Cid_etal05}) and the SSP-MILES models (Vazdekis \etal\ 2009, submitted). Figure 1a shows the mass-weighted mean ages, binned for the dynamical mass (M$_{dyn}$). The mean ages of the three subsamples coincide, except for the two central mass bins, 7.6$\times$10$^{11}$ $<$ M$_{dyn}$ $<$ 1.8$\times$10$^{12}$. The average SFHs of the subsamples in the central mass bins (Figure 1b) show that, while BCGs and Isolated ETGs have a similar evolutionary history, fossil SFHs are $\sim$ 1 Gyr younger.

\section{Discussion}

The present study agrees with previous works, which mostly found no differences between mean ages of field galaxies and either BCGs (e.g. \cite[von der Linden \etal\ 2007]{vonderLinden07}) or fossils (\cite[La Barbera \etal\ 2009]{La Barbera_etal09}). Nevertheless, fossil catalogs contain so few and heterogeneous members, that the concept of a collective behavior can be misleading. Partial results can also be relevant and our study indicates that 40 $\%$ of the fossils are younger than their BCG and isolated counterparts, suggesting that gas-rich galaxies contributed to the fossil formation. Although, fossil groups can be precursors of the isolated ETGs, the independence of ages on the environment (Figure 1a) points to an internal mechanism driving the star formation quench.

\end{document}